# First principles prediction of the Al-Li phase diagram including configurational and vibrational entropic contributions

Wei Shao [a,b], Sha Liu [b,c,*], Javier LLorca [a,b,*]

[a] Department of Materials Science. Polytechnic University of Madrid/Universidad Politécnica de Madrid. E. T. S. de Ingenieros de Caminos, 28040, Madrid, Spain
[b] IMDEA Materials Institute, C/Eric Kandel 2, Getafe, 28906, Madrid, Spain
[c] State Key Lab of Metastable Materials Science & Technology, College of Materials Science & Engineering, Yanshan University, Hebei street 438, Qinhuangdao, 066004, P. R. China.

* Corresponding authors at Department of Materials Science. Polytechnic University of Madrid. E. T. S. de Ingenieros de Caminos, 28040, Madrid, Spain and IMDEA Materials Institute, C/Eric Kandel 2, Getafe, 28906, Madrid, Spain
E-mail addresses: javier.llorca@upm.es, javier.llorca@imdea.org (J. LLorca), shaliu@ysu.edu.cn (S. Liu)

## Abstract

The whole Al-Li phase diagram is predicted from first principles calculations and statistical mechanics including the effect of configurational and vibrational entropy. The formation enthalpy of different configurations at different temperatures was accurately predicted by means of cluster expansions that were fitted from first principles calculations. The vibrational entropic contribution of each configuration was determined from the bond length *vs.* bond stiffness relationships for each type of bond and the Gibbs free energy of the different phases was obtained as a function of temperature from Monte Carlo simulations. The predicted phase diagram was in excellent agreement with the currently accepted experimental one in terms of the stable ($AlLi$, $Al_2Li_3$, $AlLi_2$, $Al_4Li_9$) and metastable ($Al_3Li$) phases, of the phase boundaries between them and of the maximum stability temperature of line compounds. In addition, it provided accurate information about the gap between $Al_3Li$ and $AlLi$ solvus lines. Finally, the influence of the vibrational entropy on the correct prediction of the phase diagram is discussed. Overall, the methodology shows that accurate phase diagrams of alloys of technological interest can be predicted from first principles calculations.

## 1. Introduction

Equilibrium phase diagrams are at the core of current strategies to design new materials with particular microstructures or to optimize existing ones through integrated computational material engineering [1]. The phase diagram provides the information about the stable phases for a given composition at a given temperature and pressure, and is one of the keystones to establish the processing-properties-structure paradigm of materials science [2].

So far, the determination of phase diagrams has been achieved through the CALPHAD (CALculation of PHAse Diagrams) methodology [3]. This strategy begins by collecting experimental information on phase equilibria and thermochemical properties of a given system, and uses appropriate mathematical models with adjustable parameters to obtain the Gibbs free energy of the different phases. This information is used to determine the stable phases as a function of composition, temperature and pressure. Notwithstanding the huge success of the CALPHAD methodology, the the predicted phase diagrams are not always as precise as required [4,5]. In particular, the limitations imposed by kinetics and metastability to experimentally determine thermodynamics and thermochemical parameters as well as the inherent approximations of the CALPHAD methodology hinder the accuracy.

Another approach to determine the phase diagram is through the combination of first principles calculations and statistical mechanics [6]. First principles calculations based on density functional theory (DFT) can be used to obtain the mixing enthalpies of different atomic configurations (possible arrangements of chemical species on a crystal lattice) at 0 K. The configurations on the convex hull in the enthalpy-composition space are ground state phases at 0 K but other configurations above the convex hull may appear at higher temperatures due to thermal fluctuations [7]. The thermodynamic properties of each phase at high temperature depends on all possible microstates but it is not feasible to determine the energies of all possible microstates by first principles calculation [8]. This limitation can be overcome by the construction of cluster expansion (CE) Hamiltonians [9,10]. The CE expresses the energy of a given configuration as a linear combination of atomic clusters multiplied by the corresponding cluster interaction coefficients, that are fitted from the mixing enthalpies obtained by DFT for a (sufficiently) large number of configurations. Thus, the mixing enthalpies of many different microstates can be accurately and efficiently predicted using the CE formalism and the free energies of different configurations (including the entropic contribution due to the configurational disorder) can be obtained as a function of temperature from Monte Carlo simulations using the principles of statistical mechanics. This information enables the parameter-free prediction of the phase diagram [11,12].

One limitation of this strategy to determine the phase diagrams is that requires previous knowledge of the crystal lattices which leads to phases in the convex hull because it is not possible to fit CE Hamiltonians for an unlimited number of lattices. This information is normally obtained from previous experimental data. In addition, the accurate prediction of the phase diagram also demands to account for the contribution of vibrational disorder - on top of the configurational one - to the free energy [13,14]. Even though the vibrational entropy contribution is normally much smaller than the configurational one, it cannot be ignored in many cases to make accurate predictions of the stable phases as a function of temperature, the phase boundaries or the solubility [15-18]. A typical example is the stability of $\theta$ and $\theta'$ phases in the Al-Cu phase diagram. Both phases have the same stoichiometry ($Al_2Cu$) but different lattices. $\theta'$ is in the convex hull and remains the stable phase up to 550K, but becomes metastable (and $\theta$ becomes the stable phase) at higher temperature due to the contribution of the vibrational disorder to the free energy [9], and the solvus line of $\theta'$ also changes accordingly.

The standard strategy to account for the vibrational entropy contribution to the free energy is based in the quasi-harmonic approximation [19]. Within this framework, the thermal properties of solid materials are traced back to those of a system of non-interacting phonons whose frequencies are, however, allowed to depend on volume or on other thermodynamic constraints. Thus, phonon calculations have to be carried out to determine the force constant matrix of each atom, which reflects the energy response from lattice vibration at each temperature [19]. Unfortunately, phonon calculations are very expensive computationally [20,21], and it is not possible to conduct phonon calculations for all possible microstates. This effort can be, however, dramatically reduced in systems with simple lattices because the force constant matrix can be transferred among different configurations and the stretching and bending stiffness of the different bonds can be accurately approximated through a bond length *vs*. bond stiffness relationship (thereafter denominated L-S relationship) in the force constant matrix [22-26]. Thus, the L-S relationship is determined for several configurations by comparison with full phonon calculations, and it is used to predict the force constant matrix of atoms in other configurations with the same lattice structure with minimum computational effort. This methodology has been successfully used to calculate the vibrational entropy contribution to the free energy of different intermetallic compounds [22-24] but - to the authors' knowledge - it has never been used to predict the whole phase diagram including several phases.

In this investigation, the Al-Li phase diagram is predicted by first principles calculations and statistical mechanics principles including the effect of vibrational entropy. To this end, the L-S relationships of nearest-neighbor bonds in the Al-Li system were obtained from several ordered configurations, and the accuracy of this approach to calculate the phonon density of states was proven by comparison with the full phonon calculations. Then, the formation enthalpies of many configurations were determined using first principles calculations and the vibrational free energy of each configuration was included using the L-S relationship. Temperature-dependent CEs were fitted taking into account the formation enthalpy and vibrational free energy of each configuration, and they were used to predict the free energies of the different phases. The whole Al-Li phase diagram was predicted from this information and compared with the currently established phase diagram [27,28]. In addition, the effect of vibrational entropy was assessed by comparison with our previous predictions of the Al-Li phase diagram without this contribution [29].

## 2. Methodology

### 2.1. Bond length *vs.* bond stiffness relationship

In Al-Li system, the stable phases share an either fcc-based or bcc-based ordered structure. Thus, Al (space group $Fm\bar{3}m$) and $Al_3Li$ (space group $Pm\bar{3}m$) are ordered structures on the fcc lattice, while the AlLi (space group $Fd\bar{3}m$), $Al_2Li_3$ (space group $R\bar{3}m$), $AlLi_2$ (space group $Cmcm$), $Al_4Li_9$ (space group $C2/m$) and Li (space group $Im\bar{3}m$) are ordered structures on the bcc lattice. The L-S relationship is related to the lattice symmetry [30] and only face-centered cubic (fcc) and body-centered cubic (bcc) lattices need to be considered because all phases in the Al-Li system are either fcc or bcc. Four ordered configurations were used to determine the L-S relationship for the fcc lattice and only three for the bcc lattice. The structures of the selected configurations were fully relaxed at P=0 using first principles calculations with Quantum Espresso [31]. The electron exchange-correlation was described using the generalized gradient approximation with the Perdew-Burke-Ernzerhof exchange-correlation functional and ultrasoft pseudo potentials [32]. The cut-off energy for the plane waves was 114 Ry. The Monkhorst-Pack scheme [33] was used for k-point sampling with a density of 40 points/Å$^{-1}$. They were compressed and expanded up to 10% in volume to create cells with different bond lengths. For each volume, a supercell of 12×12×12 unit cells was used and a displacement of 0.2 Å was applied for each symmetrically different atom in the unit cell.

The force constant matrix, $\Psi_{i\alpha,j\beta}$ (where $i$ and $j$ stands for the atoms and $\alpha$ and $\beta$ for the Cartesian directions) can be calculated from the forces on atom $j$, $F_{j\beta}$ due to the $u_{i\alpha}$ displacement of atom $i$ in direction $\alpha$ from full phonon calculations based on supercells according to [34],

$$\Psi_{i\alpha,j\beta} = \frac{\partial F_{j\beta}}{\partial u_{i\alpha}} = \frac{\partial^2 E}{\partial u_{i\alpha}\partial u_{j\beta}} \tag{1}$$

where $E$ is the energy of the perturbed supercell. $\Psi_{i\alpha,j\beta}$ between symmetrically different pairs of atoms $i$ and $j$ can be expressed as $\Psi(i,j)$, and is usually simplified as a diagonal matrix with only two independent stiffness terms, namely the stretching stiffness ($s$) and the bending stiffness ($b$) [35, 36], according to

$$\Psi(i,j) = \begin{pmatrix} b & 0 & 0 \\ 0 & b & 0 \\ 0 & 0 & s \end{pmatrix} \tag{2}$$

A L-S relationship for each type of bonding interaction (either Al-Al, Li- Li or Al-Li) was fitted from the values $b$ and $s$ obtained for each lattice deformed in tension and compression up to 10%.

## 2.2. Vibrational free energy

Following the quasi-harmonic approximation, the vibrational free energy of a configuration is determined by Born-von Karman model according to [37]

$$F_{vib}(V,T) = E(V) - TS_{vib}(V,T) \tag{3}$$

where $E(V)$ is the volume-dependent energy of the configuration, $T$ the absolute temperature and $S_{vib}(V,T)$ the vibrational entropy. $E(V)$ and $S_{vib}(V,T)$ at a given volume can be expressed as [38, 39]

$$E(V) = \int \hbar\omega \, n(\omega) \left[ \frac{1}{2} + \frac{\vartheta}{1+\vartheta} \right] d\omega \tag{4}$$

$$S_{vib}(V,T) = -k_B \int \frac{n(\omega)}{1-\vartheta} \left[ \vartheta log(\vartheta) + (1-\vartheta)log(\vartheta) \right] d\omega \tag{5}$$

where $\hbar$ is the reduced Planck's constant, $k_B$ the Boltzmann constant, $\vartheta = e^{-\hbar\omega/k_BT}$ and $n(\omega)$ represents the phonon density of states (DOS), which is defined as the total number of modes with frequencies between $\omega$ and $\omega + d\omega$ per unit volume. For each configuration, the phonon DOS is determined from the dynamical matrix $D_{i,j}(q)$ which is obtained by means of the Fourier transform of $\Psi(i,j)$ at wavevector $q$ from which the eigenfrequencies $\omega$ are computed. After all eigenfrequencies are computed, the phonon DOS is calculated [9].

The $\Psi(i,j)$ matrix can be obtained either from the full phonon calculation (as indicated in section 2.1) or through the L-S relationship as long as the equilibrium bond lengths are available. To this end, 154 different configurations in the fcc lattice and 489 configurations in the bcc lattice were generated. They were fully relaxed at P=0 using Quantum Espresso with the same parameters used for the ordered configurations above. The lattice distortion of each configuration after relaxation was determined to assess whether the lattice symmetry changed. Following the criteria reported in the literature [40,41], only configurations whose distortion was below 10% were considered to keep the original lattice symmetry and used to fit the CE for each lattice [29]. In fact, 18 configurations of the fcc system and 297 configurations of the bcc system changed the original lattice during relaxation and they were no longer considered. The equilibrium bond lengths were obtained for each stable configuration and the $\Psi(i,j)$ matrices in each configuration were predicted through the L-S relationship. This information was used to determine $F_{vib}(V,T)$ for each configuration through eqs. (3) to (5).

## 2.3. Temperature dependent cluster expansion

The temperature-dependent mixing enthalpy $H^{mix}$ of a configuration with composition $Al_{1-x}Li_x$ - including the vibrational entropy contribution - can be expressed as [42]

$$H^{mix}_{Al_{1-x}Li_x}(T) = F^{Al_{1-x}Li_x}_{vib}(T) - (1-x)F^{Al}_{vib,s}(T) - xF^{Li}_{vib,s}(T) \qquad (6)$$

where $F^{Al}_{vib,s}(T)$ and $F^{Li}_{vib,s}(T)$ stand for the vibrational free energy of pure Al and Li in the corresponding lattice at temperature $T$. It should be noted that the mixing enthalpy given by eq. (6) will be used to fit the CE but it does not correspond to the actual formation enthalpy, $H^f$, that is expressed as

$$H^f_{Al_{1-x}Li_x}(T) = F^{Al_{1-x}Li_x}_{vib}(T) - (1-x)F^{Al}_{vib}(T) - xF^{Li}_{vib}(T) \qquad (7)$$

where $F^{Al}_{vib}(T)$ and $F^{Li}_{vib}(T)$ stand for vibrational free energy of pure Al and Li in their equilibrium structure, namely fcc Al and bcc Li.

The temperature-dependent CE formalisms for the fcc and bcc lattices can be fitted from the temperature-dependent mixing enthalpies of the configurations in fcc and bcc lattices, respectively. They are expressed as [43, 44]

$$H^{mix}(\vec{\sigma}, T) = \sum_\alpha m_\alpha J_\alpha(T) \langle \varphi(\vec{\sigma}) \rangle_\alpha \tag{8}$$

where the summation index stands for all distinct clusters $\alpha$ under symmetry operation. $m_\alpha$ stands for the multiplicity of clusters $\alpha$ and $J_\alpha(T)$ is the effective cluster interaction (ECI) coefficient of each cluster at temperature $T$. $\langle \varphi(\vec{\sigma}) \rangle_\alpha$ is the cluster function of $\alpha$ defined as a product of orthogonal point functions, and $\vec{\sigma}$ is vector that indicates which type of atom sits on lattice site $i$, i.e. $\sigma_i$ = -1 when the lattice site $i$ is occupied by an Al atom and $\sigma_i$ = +1 when the lattice site $i$ is occupied by a Li atom. The number of clusters and the ECI coefficients for both fcc and bcc lattices were fitted using the ATAT package [45]. It should be noted that the mixing enthalpies provided by the CE formalism can used to predict actual formation enthalpy of any configuration including the vibrational entropy contribution.

## 2.4. Phase diagram construction

The grand canonical free energy $\emptyset$ is defined as [46]

$$\beta \emptyset(\beta, \Delta\mu) = -lnZ \tag{9}$$

where $\beta = \frac{1}{k_B T}$ is the reciprocal of temperature, $\Delta\mu$ represents the difference in chemical potential between the Al and Li and the partition function $Z$ is calculated as [47]

$$Z = \sum_i exp\left(-\beta N\left(H_i^f - \Delta\mu x\right)\right) \tag{10}$$

where $N$ is the number of atoms in the crystal lattice, $H_i^f$ formation enthalpy of each possible microstate $i$ and $x$ the atomic fraction of Li in the configuration.

From eqs. (9) and (10), the grand potential $\emptyset$ for each phase can be written as

$$d(\beta\emptyset) = N(\langle H^f \rangle - \Delta\mu x)d\beta - N\beta x \, d(\Delta\mu) \tag{11}$$

where $\langle H^f \rangle$ stands for the average formation enthalpy per atom. The partition function $Z$ was evaluated by means of semi-grand canonical MC simulations for different values of $T$ and $\Delta\mu$ according to eq. (10) and then $\emptyset$ can be calculated through thermodynamic integration. For a given $\Delta\mu$, $\emptyset$ is given as [48]

$$\emptyset(\beta_1, \Delta\mu) = \emptyset(\beta_0, \Delta\mu) + \frac{N}{\beta} \int_{\beta_0}^{\beta_1} (H^f - \Delta\mu x) \, d\beta \qquad (12)$$

and as

$$\emptyset(\beta, \Delta\mu_1) = \emptyset(\beta, \Delta\mu_0) - N \int_{\Delta\mu_0}^{\Delta\mu_1} x \, d\Delta\mu \qquad (13)$$

for a given $\beta$, where $\beta_0$, $\Delta\mu_0$ and $\beta_1$ and $\Delta\mu_1$ stand for the range of temperatures and chemical potentials explored in the MC simulations.

In the case of solid-state transformations at atmospheric pressure, the difference between Gibbs and Helmholtz free energies can be neglected and the Gibbs free energy $G$ is expressed as [47]

$$G = \emptyset - \Delta\mu x \qquad (14)$$

The semi-grand canonical MC simulations [49] of fcc and bcc Al-Li systems were carried out using ATAT package in periodic supercells of dimensions 20×20×20 primitive unit cells. The temperature range varied from 10 K to 1000 K with a temperature increment of 10 K for each chemical potential. At each temperature, MC simulations were carried out from the ground state phases with increasing and decreasing $\Delta\mu$ with an increment of 0.005 eV/atom in the chemical potential. For each value of $T$ and $\Delta\mu$, a MC simulation included 2000 passes for equilibrium, followed by 5000 passes for calculating the thermodynamic averages. For two adjacent phases with the same lattice structure, the two-phase equilibrium region in the phase diagram can be determined by the intersection of $\emptyset$ obtained from increasing and decreasing $\Delta\mu$. For two adjacent phases with different lattice structures, the two-phase equilibrium region in the phase diagram can be determined by the common tangent of their Gibbs free energies curves at a given temperature.

## 3. Results

### 3.1. Bond length *vs.* bond stiffness relationship

There are three types of chemical bonds between different atomic species in Al-Li alloys, namely Al-Al, Li-Li and Al-Li bonds. Their L-S relationships in the fcc lattice were obtained from the ordered Al, $Al_3Li$, AlLi and Li compounds, whose structures are shown in Fig. 1. The first-nearest neighbor bonds in fcc Al and fcc Li (Figs. 1(a) and (d)) are Al-Al and Li-Li and their equilibrium bond lengths are 2.837 Å and 3.05 Å, respectively. The first-nearest neighbor bonds in fcc $Al_3Li$ (Fig. 1(b)) are Al-Al and Al-Li with the same equilibrium bond length of 2.828 Å. Finally, the first-nearest neighbor bonds in fcc AlLi (Fig. 1(c)) are Al-Al (2.820 Å), Li-Li (2.820 Å) and Al-Li (2.839 Å). They were also used to fit the L-S relationships to ensure that they are transferable.

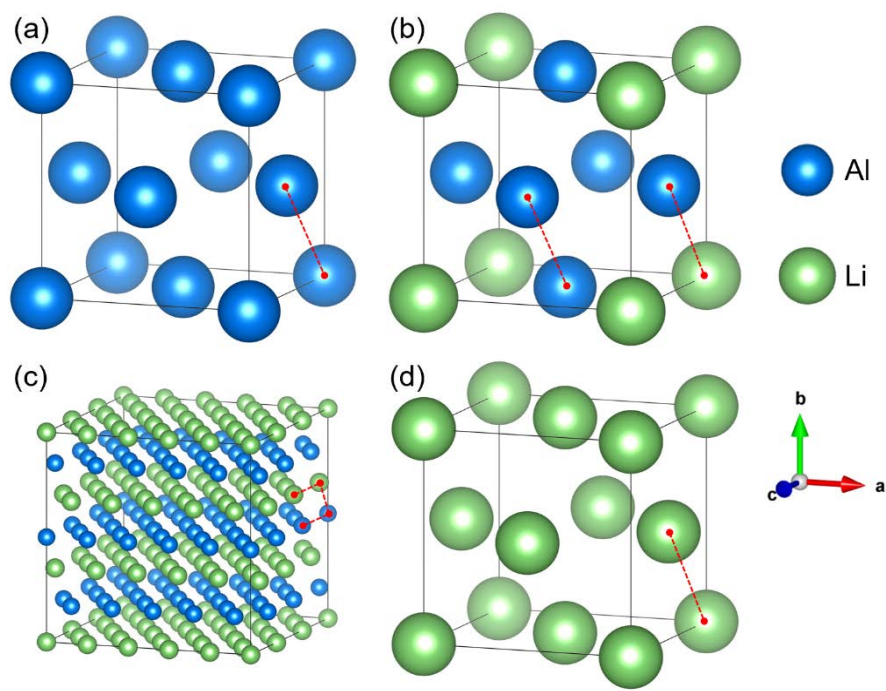

**Fig. 1.** Crystal structures of ordered Al-Li configurations with fcc lattice. (a) Al; (b) $Al_3Li$; (c) AlLi; (d) Li. The first nearest-neighbor bonds are indicated with red lines.

The force constant matrix of the three types of bonds in the ordered fcc Al-Li configurations with different volumes were determined from full phonon calculations and the corresponding L-S relationships. The stretching and bending stiffnesses of each bond are depicted in Fig. 2. The stretching stiffness decreases dramatically with bond length in the three bonds, while the bending stiffness increases slightly. The stretching and bending stiffness terms for the three types of bonds (Al-AL, Al-Li and Li-Li) are very close, regardless of the ordered configuration, and this result indicates that the force constant matrix is transferable for different configurations with the same lattice. The L-S relationships of stretching and bending stiffness for each type of bond were fitted by quadratic polynomials using the least-squares method and they are plotted as solid lines in Fig. 2. They can be used to predict the force constant matrix of configurations with many atoms arranged in a fcc lattice.

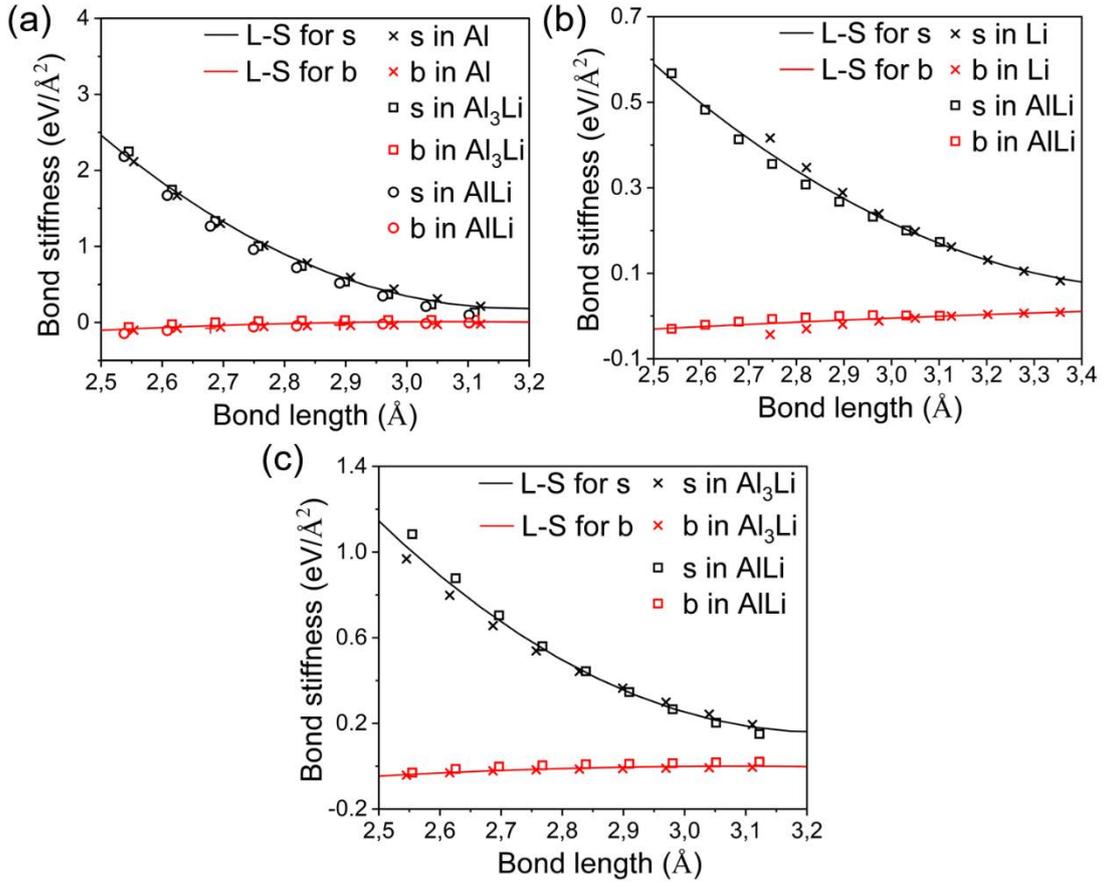

**Fig. 2.** Bond stiffness *vs.* bond length relationships fitted from ordered fcc Al-Li configurations. The symbols stand for the results obtained from full phonon calculations according to eq. (1). The solid lines are the fitted polynomials. (a) Al-Al bond; (b) Li-Li bond; (c) Al-Li bond. *s* stands for the stretching stiffness and *b* for the bending stiffness.

The Al, AlLi and Li ordered configurations, whose structures are displayed in Fig. 3, were used to determine the L-S relationships in the bcc lattice. The first-nearest neighbor bonds in bcc Al and Li are Al-Al and Li-Li (Figs. 3(a) and (c)) and their equilibrium bond lengths are 2.778 Å and 2.968 Å, respectively. The first-nearest neighbor bonds in bcc AlLi (Fig. 1(b)) are Al-Al, Al-Li and Li-Li with the same equilibrium bond length of 2.736 Å in all cases.

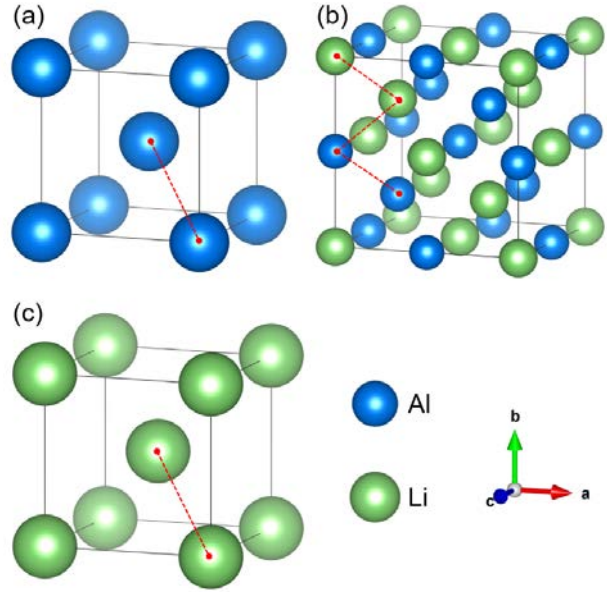

**Fig. 3.** Crystal structure of ordered Al-Li configurations with bcc lattice. (a) Al; (b) AlLi; (c) Li. The first nearest-neighbor bonds are indicated with red lines.

The force constant matrix of the three types of bonds in the ordered bcc Al-Li configurations with different volumes were determined from full phonon calculations. The corresponding L-S relationships are depicted in Fig. 4. The stretching and bending stiffnesses in the three bonds follow the same trends found previously in the fcc ordered configurations (Fig. 2). The L-S relationships of stretching and bending stiffness for each type of bond were fitted into quadratic polynomials using the least-squares method and they are plotted as solid lines in Fig. 4. They were used to predict the force constant matrix of configurations with the bcc lattice.

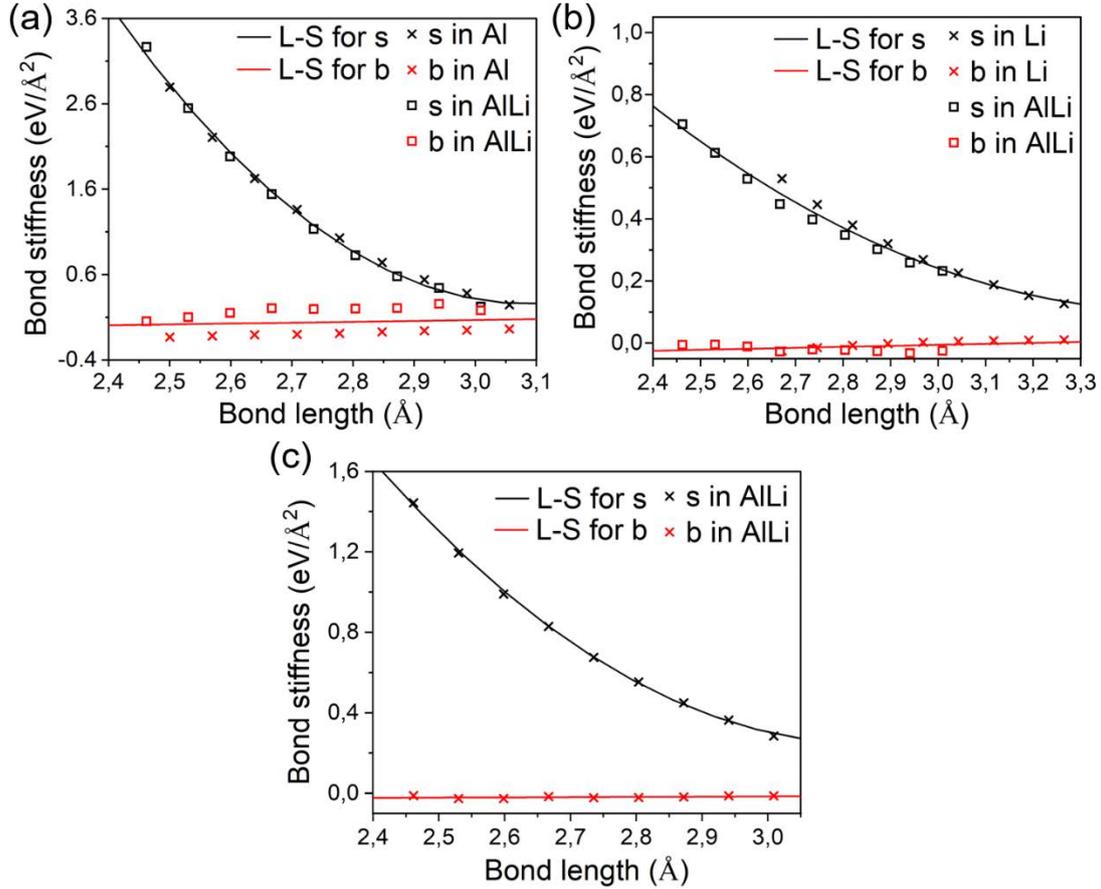

**Fig. 4.** Bond stiffness *vs.* bond length relationships fitted from ordered bcc Al-Li configurations. The symbols stand for the results obtained from full phonon calculations according to eq. (1). The solid lines are the fitted polynomials. (a) Al-Al bond; (b) Li-Li bond; (c) Al-Li bond. *s* stands for the stretching stiffness and *b* for the bending stiffness.

### 3.2. Vibrational free energy

The accuracy of the L-S relationship to obtain the phonon DOS and the vibrational entropy contribution to the Gibbs free energy can be assessed by comparison with the results obtained from full phonon calculations for the most relevant phases in the Al-Li system. The vibrational free energies, $F_{vib}$, of Al, Al$_3$Li, AlLi, Al$_2$Li$_3$, Al$_4$Li$_9$ and Li calculated using both approaches are plotted as a function of temperature in Figs. 5(a) to 5(f), and the corresponding phonon DOS are inserted in each figure. The red curves stand for the results obtained from full phonon calculations while the black curves indicate the vibrational free energy and phonon DOS predicted with the L-S relationships. The phonon DOS predicted with the L-S relationships are consistent with those obtained from full phonon calculations for all the different intermetallic compounds with fcc and bcc lattices. As a result, the vibrational entropic contribution - that is obtained by the integration of the phonon DOS through eq. (5) - obtained with the L-S relationships is very close to the actual one determined from the full phonon calculations. The small differences are similar for all phases and are not significant from the viewpoint of accuracy in the formation enthalpy.

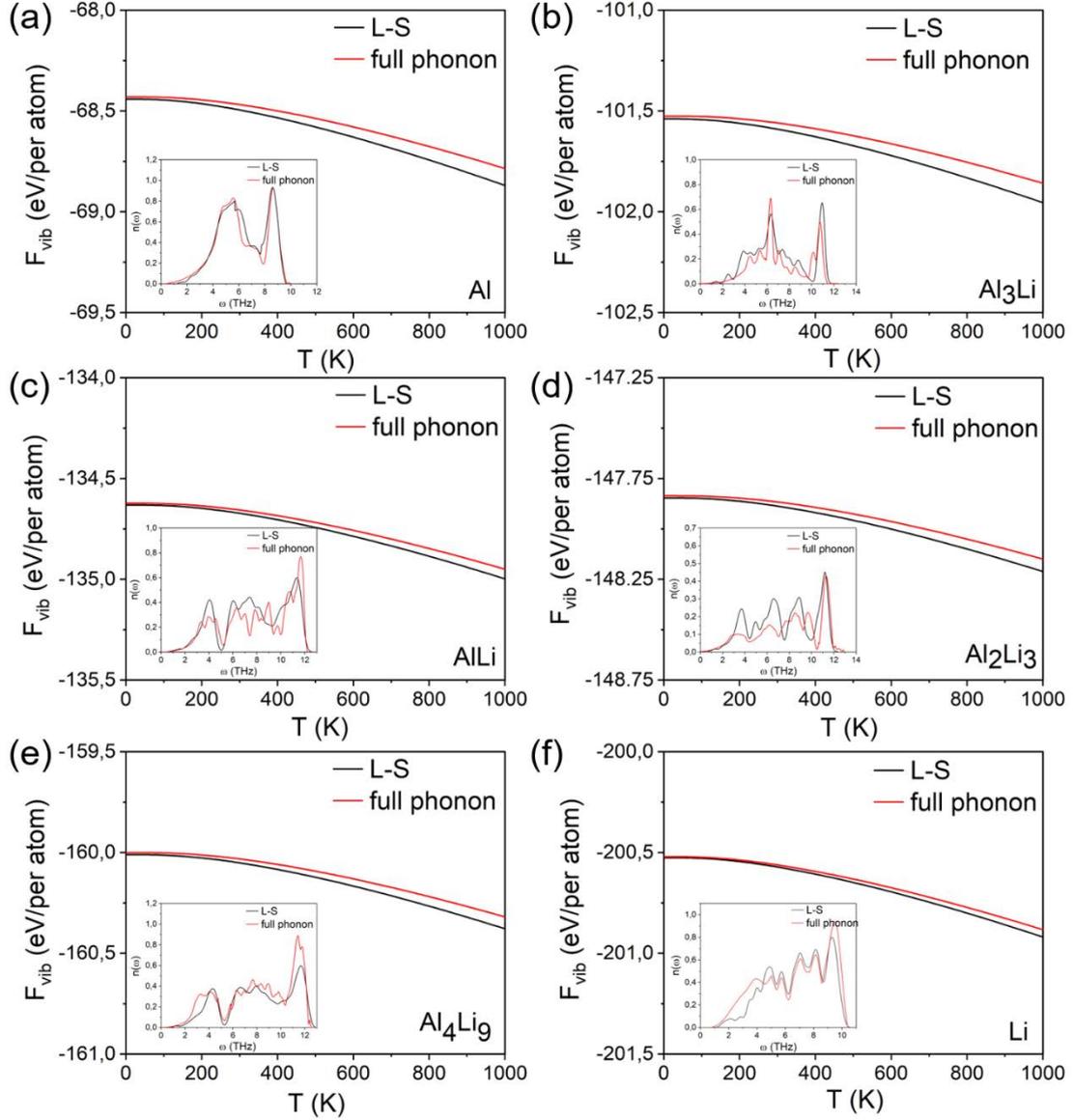

**Fig. 5.** Vibrational free energy of phases in Al-Li system determined by full phonon calculations and predicted with the L-S relationships. (a) Al; (b) $Al_3Li$; (c) AlLi; (d) $Al_2Li_3$; (e) $Al_4Li_9$; (f) Li.

### 3.3. Temperature-dependent cluster expansions and ground state phases

Temperature-dependent CEs for fcc and bcc Al-Li systems were fitted and the ECI coefficients as well as the cluster features are listed in Tables S1 and S2 in the Supplementary Materials. The formation enthalpies, $H^f$, of different configurations determined from the combination of first principles calculations and the L-S relationships and predicted with the CEs are compared in Fig. 6 for three different temperatures: 0 K, 500 K and 1000 K. The red crosses and open symbols stand for the bcc configurations calculated using both strategies while the blue crosses and open symbols provide the same information for the fcc configurations. The formation enthalpies at other temperatures are plotted in Fig. S1 in the Supplementary Materials. The $H^f$ predicted with the CE formalism are in good agreement with those obtained from first principles calculations and the L-S relationships. The cross-validation scores for the fcc and bcc configurations were 0.0067 eV/atom and 0.0148 eV/atom, respectively.

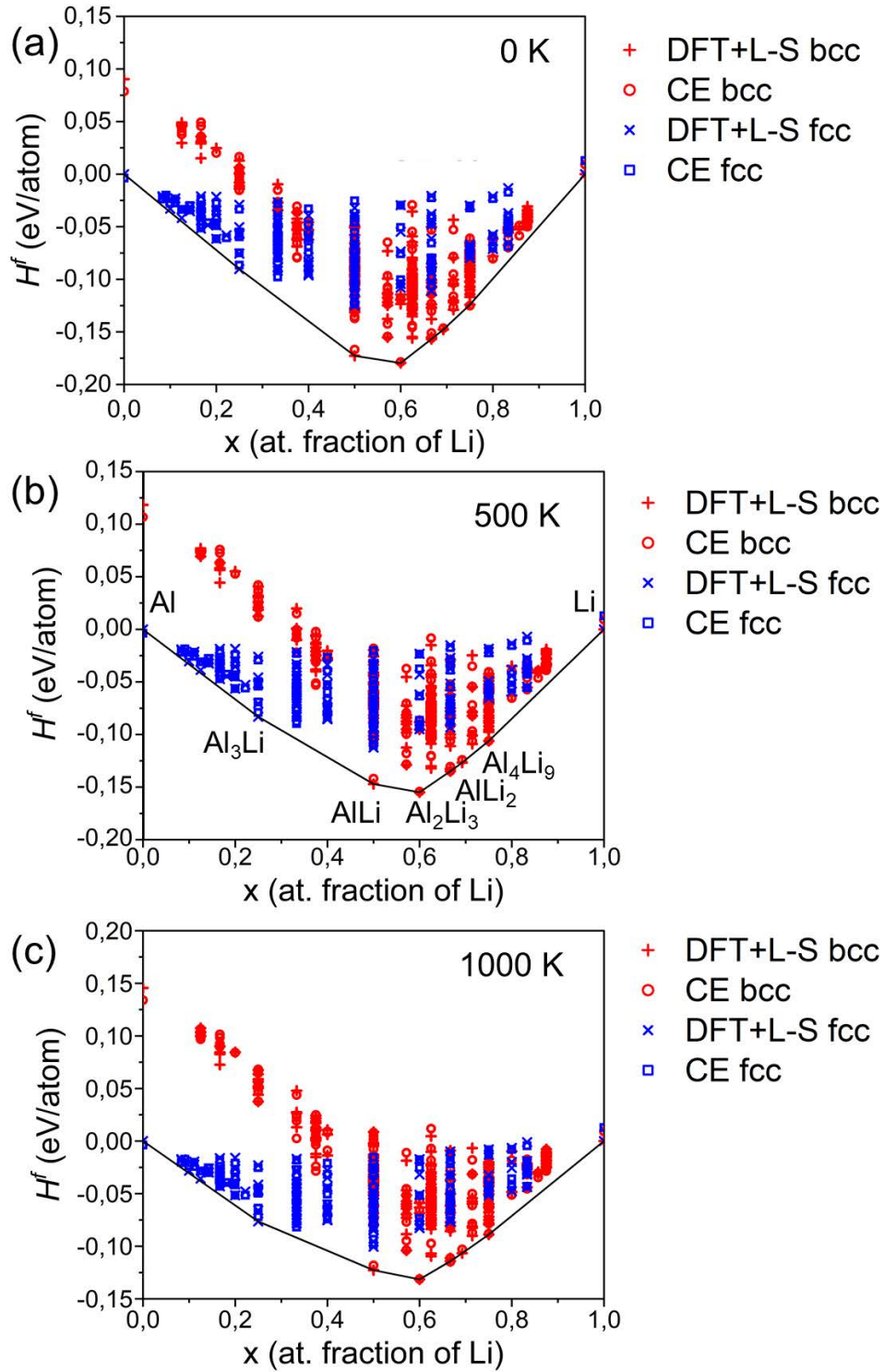

**Fig. 6.** Formation enthalpies $H^f$ of different configurations in the Al-Li system at different temperatures. (a) 0 K; (b) 500K; (c) 1000K. Crosses stand for the enthalpies calculated using first principles calculations and the L-S relationships while open symbols provide the results given by the CE formalism. Red symbols stand for bcc configurations and blue symbols for fcc.

The ground state phases located on the convex hull at 0 K can be found in Fig. 6a. They are Al, Al$_3$Li ($\delta'$), AlLi ($\delta$), Al$_2$Li$_3$, AlLi$_2$, Al$_4$Li$_9$ and Li. Al and Al$_3$Li ($\delta'$) have fcc structures while the others are bcc. At higher temperature, $H^f$ provides information about the stability of the phases including the contribution from vibrational entropy and the same phases are found in convex hull at 500 K (Fig. 6(b)) and 1000 K (Fig 6(c)). Al$_3$Li ($\delta'$) is located almost on the tie-line connecting Al and AlLi ($\delta$) at 0K, and its formation enthalpy is only 4.17 meV/atom lower than that of the straight line connecting Al and AlLi, which is the same order of the accuracy in first principles calculations. Thus, it is difficult to establish whether Al$_3$Li ($\delta'$) is ground state phase at 0 K. Nevertheless, the formation enthalpy of Al$_3$Li ($\delta'$) with respect to the straight line connecting Al and AlLi is slightly increased with temperature up to 6.172 meV/atom at 500 K (Fig. 6(b)), and to 10.942 meV/atom at 1000 K (Fig. 6(c)).

## 3.4. Phase diagram

### 3.4.1. Al/Al$_3$Li phase boundary

The semi-grand canonical MC calculations were performed from each ground state phase in the direction of both increasing and decreasing $\Delta\mu$. In the Al-rich part, Al and Al$_3$Li are the two ground state phases which share the same fcc lattice (Fig. 6). Their phase boundaries can be determined by the grand potential $\emptyset$, which was obtained by thermodynamic integration from $\Delta\mu$ and T, as indicated in Section 2.4.

The grand potentials $\emptyset$ are plotted as a function of $\Delta\mu$ at T=500 K in Fig. 7(a) as an example. The black curve is calculated from Al by increasing $\Delta\mu$, while the red one is calculated from Al$_3$Li by decreasing $\Delta\mu$. The intersection $\emptyset_{Al} = \emptyset_{\delta'}$ corresponds to $\Delta\mu$ where Al and $\delta'$ coexist. On the left of the intersection, $\emptyset_{Al} < \emptyset_{\delta'}$, which indicates that Al is the stable phase in this $\Delta\mu$ range, while $\emptyset_{\delta'} < \emptyset_{Al}$ on the right of the intersection, and Al$_3$Li is the stable phase. The two-phase equilibrium region of Al and Al$_3$Li, expressed by $\emptyset_{Al} = \emptyset_{\delta'}$, can be mapped on to the conjugated relationship between composition $x$ and chemical potential $\Delta\mu$, as shown in Fig. 7(b), from which the composition of Al and Al$_3$Li at equilibrium can be determined. Thus, the phase boundary between Al and Al₃Li at 500K can be constructed and is marked by the dashed line in Fig. 7(c). Following the same methodology, the phase boundary between Al and Al₃Li can be determined as a function of temperature and the phase diagram in this region is constructed, as shown in Fig. 7(c). The two-phase equilibrium region of Al and Al₃Li exists up to T = 680 K. The $\emptyset - \Delta\mu$ and $x - \Delta\mu$ curves obtained by increasing $\Delta\mu$ and decreasing $\Delta\mu$ overlap when the temperature is higher than 680 K, which means Al and Al₃Li cannot be distinguished from a thermodynamic viewpoint.

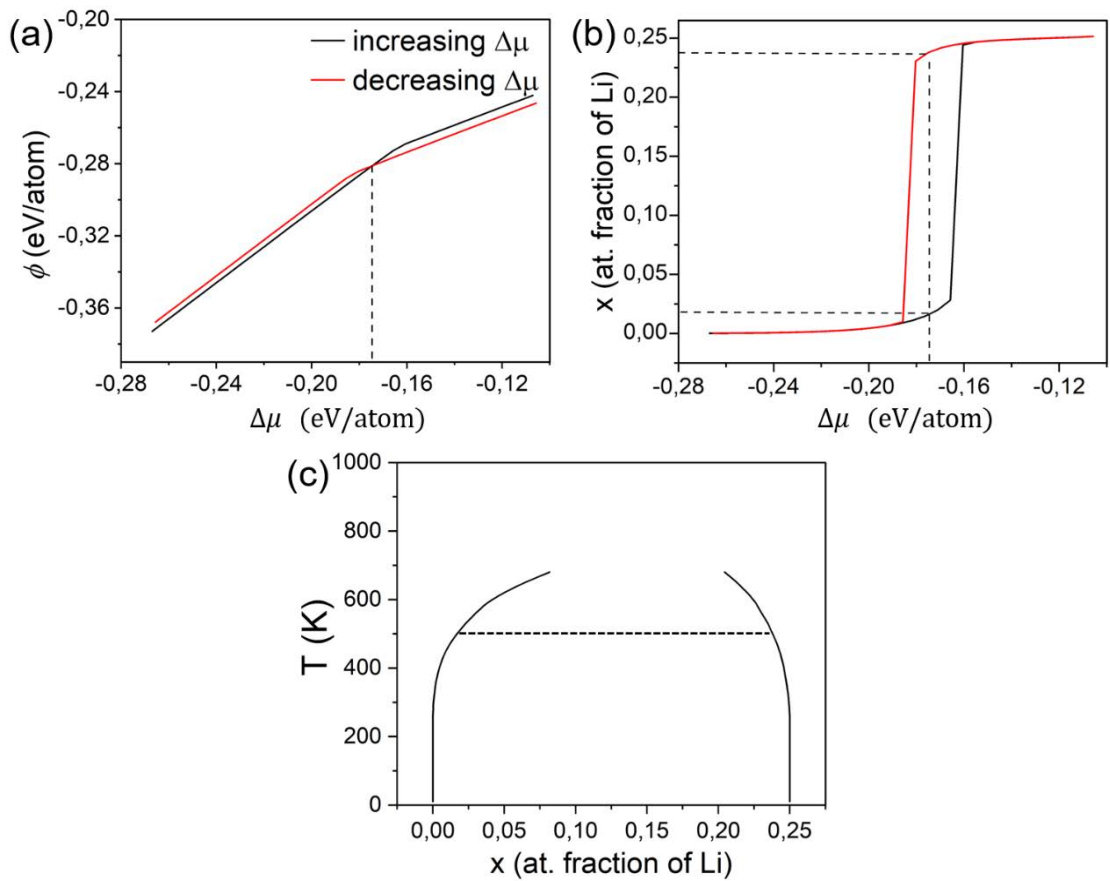

**Fig. 7.** (a) Grand potentials $\emptyset$ of Al and Al₃Li as a function of chemical potential $\Delta\mu$ at 500 K. (b) Li content (expressed by $x$) as a function of $\Delta\mu$ at 500 K. (c) Phase boundary between Al and Al₃Li.

### 3.4.2. Al₃Li/AlLi phase boundary

The other adjacent phase to Al$_3$Li is AlLi (Fig. 6), which has a bcc lattice structure. Therefore, the phase boundary between AlLi and either Al or Al$_3$Li can only be determined by the common tangent to their Gibbs free energy $G$, which were calculated by according to eq. (14). Since the Gibbs free energy of Al$_3$Li is almost on the tie-line connecting Al and AlLi at 0 K (Fig. 6(a)), it is difficult to establish whether Al$_3$Li ($\delta'$) is ground state phase at 0 K, and the Gibbs free energy of Al$_3$Li is also compared. The dash lines in Fig. 8(a) stand for the common tangent to the Gibbs free energies of Al and AlLi at each temperature, and they indicate the chemical potential at equilibrium between Al and AlLi. The Gibbs free energies of Al$_3$Li are always above the common tangent to the Gibbs free energies of Al and AlLi at each temperature, and this result indicates the Al$_3$Li is a metastable phase, mainly due to the configurational entropic contribution to the free energy because the vibrational entropic contribution enhanced the stability of Al$_3$Li.

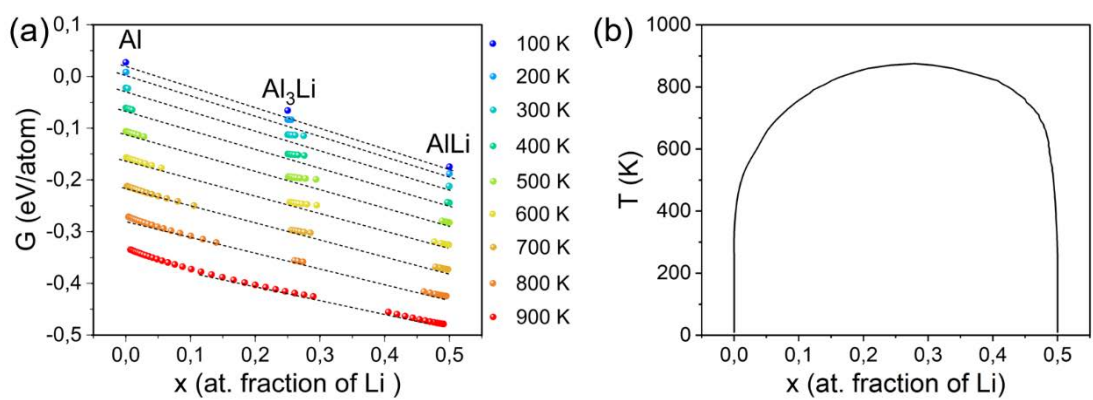

**Fig. 8** (a) Gibbs free energy of Al, Al$_3$Li and AlLi phases as a function of the Li content (expressed by $x$) at different temperatures; (b) Phase boundary between Al and AlLi.

The composition of Al and AlLi at equilibrium can be obtained for each temperature from the common tangent at the Gibbs free energy curves, and the phase boundary between them is constructed (Fig. 8(b)). The two-phase equilibrium region of Al and AlLi exist up to T = 850 K, in good agreement with experimental results [27, 28]. The Gibbs free energies of Al and AlLi overlap at $T$ > 850 K (Fig. 8(a)), and they become solid solutions above this temperature.

### 3.4.3. AlLi/Al$_2$Li$_3$ and Al$_2$Li$_3$/AlLi$_2$ phase boundaries

AlLi and Al₂Li₃ share the bcc lattice structure, and their phase boundary was determined from the thermodynamic grand potentials $\phi$, as in the case of Al and Al₃Li. Their grand potentials are plotted as a function of $\Delta\mu$ at T = 600 K in Fig. 9(a) as another example. The black curve is obtained from AlLi by increasing $\Delta\mu$, while the red one is obtained from Al₂Li₃ by decreasing $\Delta\mu$. On the left of the intersection, $\phi_{AlLi} < \phi_{Al_2Li_3}$, which indicates the Al₂Li₃ is the stable phase in this $\Delta\mu$ range, while $\phi_{AlLi} > \phi_{Al_2Li_3}$ on the right of the intersection, and Al₂Li₃ is the stable phase. The intersection of the grand potentials can be mapped on to the conjugated relationship between composition $x$ and $\Delta\mu$, as shown in Fig. 9(b), which determines the composition of AlLi and Al₂Li₃ at equilibrium. Thus, the phase boundary between AlLi and Al₂Li₃ at T = 600 K can be constructed, and is indicated by the dashed line in Fig. 9(c). Following this strategy, the phase boundary between both phases was built as a function $T$ and is plotted in Fig. 9(c).

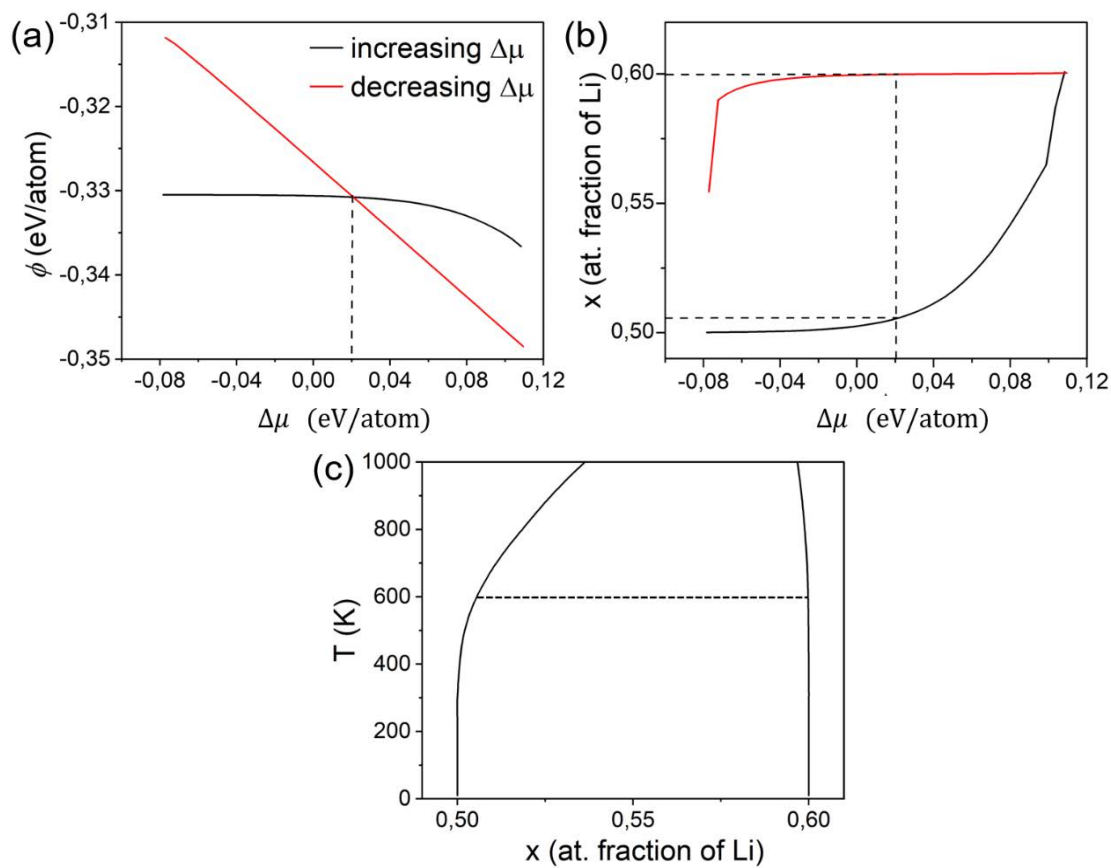

**Fig. 9.** (a) Grand potentials ∅ of AlLi and Al₂Li₃ as a function of the chemical potential Δμ at 600 K. (b) Li content (expressed by *x*) as a function of Δμ at 600 K. (c) Phase boundary between AlLi and Al₂Li₃.

The strategy to obtain the phase boundaries between Al₂Li₃ and AlLi₂ is identical to the one presented above for AlLi and Al₂Li₃. The ascending and descending curves of the thermodynamic grand potentials ∅ as a function of Δμ at T = 500 K are plotted in Fig. 10(a). The composition of Al₂Li₃ and AlLi₂ at equilibrium could be determined from their intersection (Fig. 10(b)) and this construction is repeated at different temperatures to plot the phase boundaries, which are plotted in Fig. 10(c).

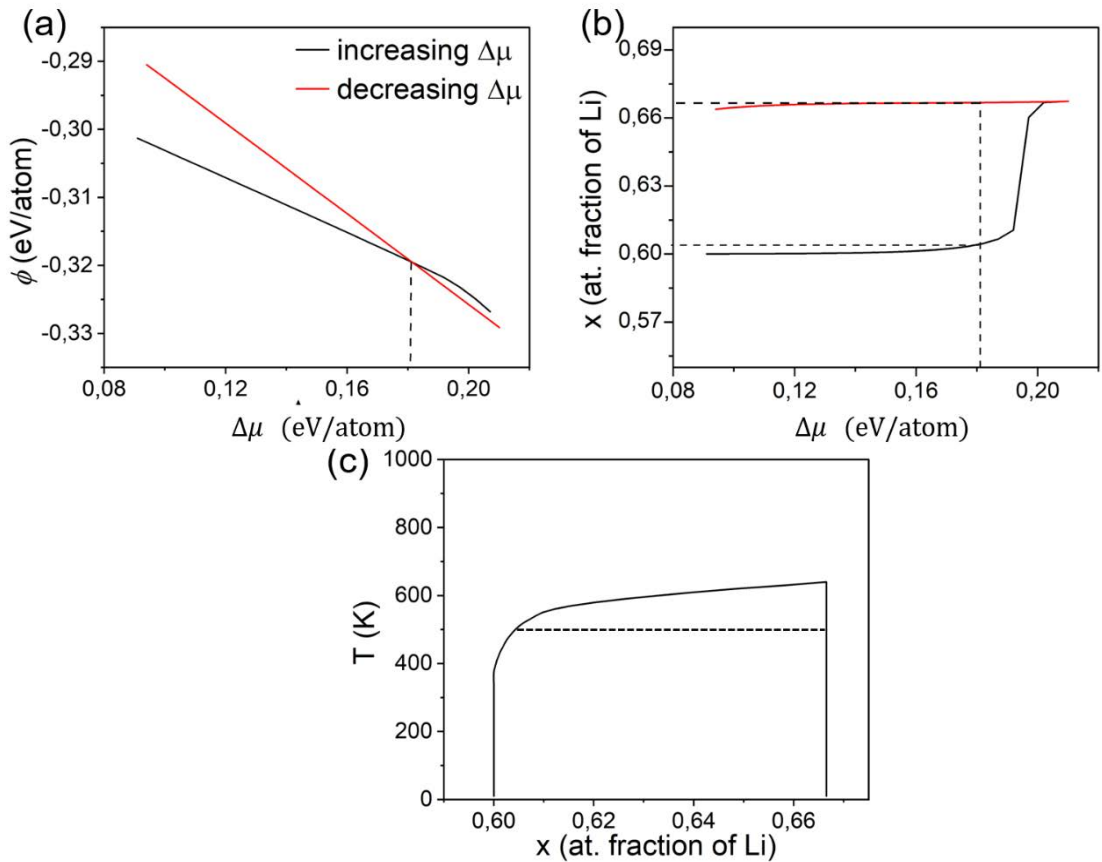

**Fig. 10.** (a) Grand potentials ∅ of Al₂Li₃ and AlLi₂ as a function of the chemical potential Δμ at 500 K. (b) Li content (expressed by *x*) as a function of Δμ at 500 K. (c) Phase boundary between Al₂Li₃ and AlLi₂.

### 3.4.4. AlLi₂/Al₄Li₉ and Al₄Li₉/Li phase boundaries

The phase boundaries between AlLi₂ and Al₄Li₉ were also determined from their thermodynamic grand potentials ∅ and they are plotted in Fig.11(a) as a function of Δμ at T = 450 K as an example. Again, the intersection between the ascending and descending grand potentials was mapped on to the conjugated relationship between Li content (expressed by $x$) and chemical potential $\Delta\mu$, as shown in Fig. 11(b). Thus, the composition of AlLi$_2$ and Al$_4$Li$_9$ at equilibrium was determined and the phase boundary between AlLi$_2$ and Al$_4$Li$_9$ at T = 450 K was obtained and is plotted with the dashed line in Fig. 11(c). The phase boundary in this region was constructed from the grand potentials at each temperature and is shown in Fig. 11(c), from which the AlLi$_2$ and Al$_4$Li$_9$ are both line compounds. The two-phase equilibrium region of AlLi$_2$ and Al$_4$Li$_9$ exists up to T = 520 K, and Al$_2$Li$_3$ and AlLi$_2$ cannot be distinguished above 520 K.

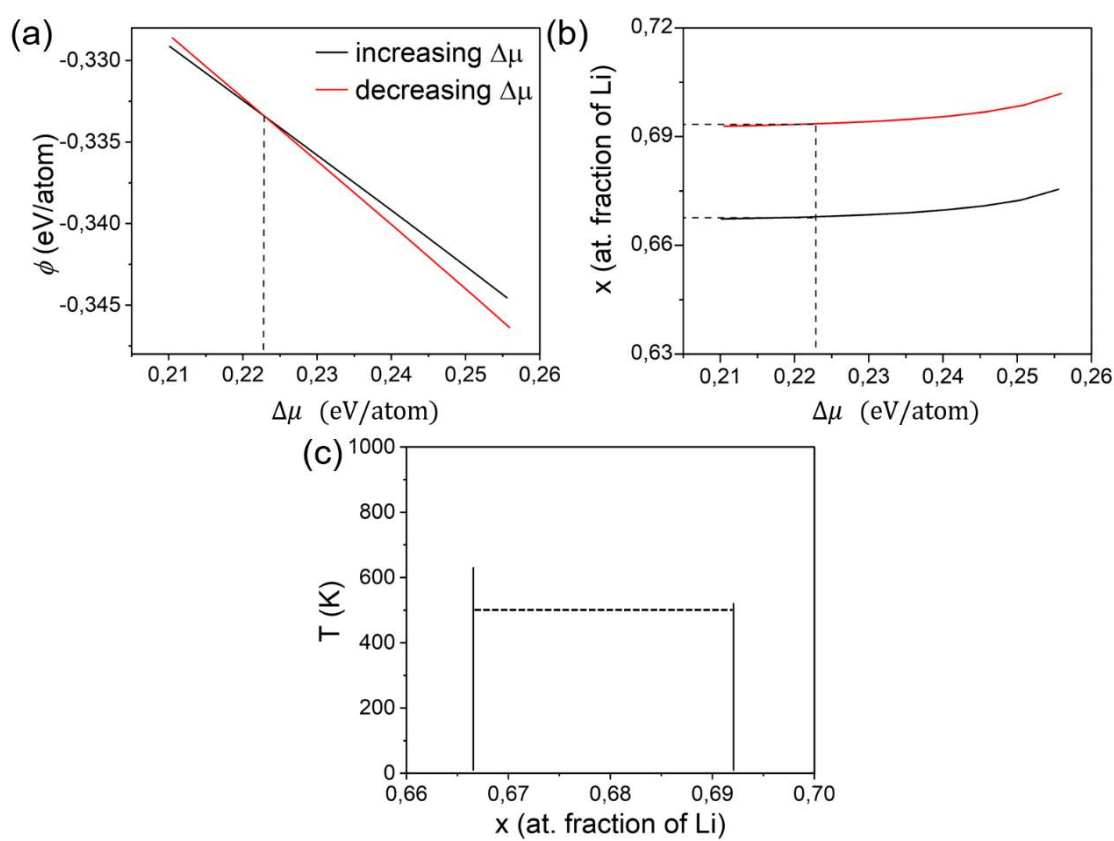

**Fig. 11.** (a) Grand potentials $\emptyset$ of AlLi$_2$ and Al$_4$Li$_9$ as a function of the chemical potential $\Delta\mu$ at 450 K. (b) Li content (expressed by $x$) as a function of $\Delta\mu$ at 450 K. (c) Phase diagram between AlLi$_2$ and Al$_4$Li$_9$.

Exactly the same construction was used to obtain the phase boundaries between Al$_4$Li$_9$ and Li. The corresponding ascending and descending $\emptyset$ *vs.* $\Delta\mu$ curves (Fig. 12(a)), the composition at equilibrium of both phases at T = 400 K (Fig. 12(b)) and the phase boundaries as a function of temperature (Fig. 12(c)) were obtained. are shown in Fig.12(a). The two-phase equilibrium region of Al$_4$Li$_9$ and Li exists up to T=520 K.

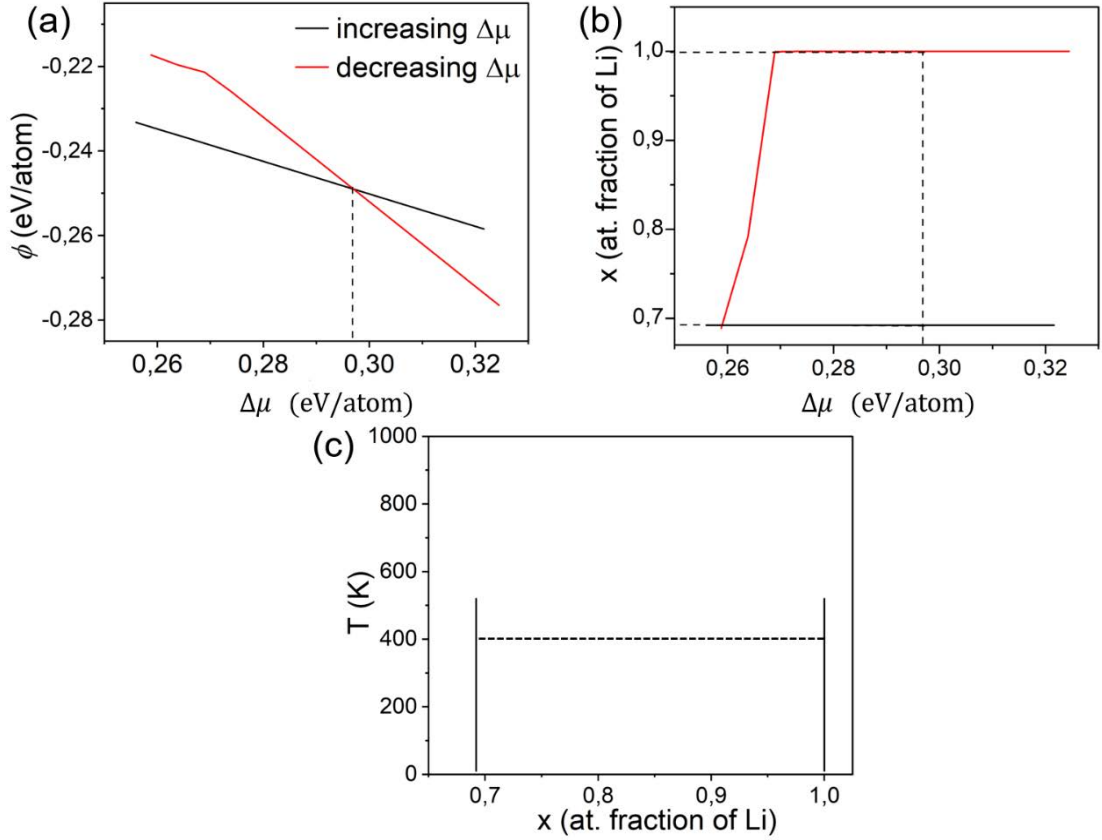

**Fig. 12.** (a) Thermodynamics grand potentials $\emptyset$ of $Al_4Li_9$ and Li as a function of the chemical potential $\Delta\mu$ at 400 K. (b) Li content (expressed by $x$) as a function of $\Delta\mu$ at 400 K. (c) Phase diagram between $Al_4Li_9$ and Li.

## 4. Discussion

The phase boundaries between each two adjacent phases were integrated to build the Al-Li phase diagram which is plotted in Fig. 13(a) together with the previous one that was obtained including only the configurational entropy contribution to the free energy [29]. $Al_3Li$ is considered as a stable phase in the previous phase diagram that only includes configurational disorder and, thus, there are two Al-$Al_3Li$ and $Al_3Li$-Al two-phase equilibrium regions (in red). However, since $Al_3Li$ is a metastable phase in the phase diagram including both vibrational and configurational entropy, the $Al_3Li$-AlLi two-phase equilibrium region has disappeared and only one Al-AlLi two-phase equilibrium region exists. Moreover, the critical temperature of Al-$Al_3Li$ two-phase region (maximum of the red dashed line) was 730 K in the phase diagram with only configurational disorder, but this value is higher than the experimental ones (620 K) [28]. This discrepancy could be expected because neglecting the vibrational contribution often leads to overestimate of phase transition temperatures [50, 51]. The phase diagram including both vibrational and configurational contributions leads to a wider two-phase equilibrium region between Al and $Al_3Li$ (black dashed lines) than the one without considering vibrational effect, and the critical temperature also decreases to 670 K, which is in better agreement with the experimental data.

$AlLi_2$ and $Al_4Li_9$ appear as line compounds in the phase diagrams with or without vibrational entropy. Nevertheless, $AlLi_2$ and $Al_4Li_9$ are only stable up to 300 K and 500 K, respectively, when vibrational disorder is not included. They reach 640 K and 520 K, respectively, when vibrational entropy is taken into account, which is much closer to the experimental data (600 K and 545 K, respectively). The solubilities of AlLi and $Al_2Li_3$ are also influenced by the vibrational entropy. The Li content in the two phase AlLi- $Al_2Li_3$ region reaches 0.6 at 700 K if only configurational disorder is included but is dramatically reduced at this temperature if vibrational entropy is taken into account. In addition, $Al_2Li_3$ is a line compound if only configurational disorder is included but it forms a two-phase region with $AlLi_2$ if vibrational entropy is considered.

The phase diagram calculated taking into account both vibrational and configurational entropic contributions is also compared with the accepted experimental phase diagram [27, 28] in Fig. 13(b). Both phase diagrams predict that $Al_3Li$ is a metastable phase. The Al-AlLi two-phase equilibrium regions in the both phase diagrams are almost superposed and a small difference only appears at high temperature. The phase boundary between AlLi and $Al_2Li_3$ is also accurately predicted in the phase diagram. Moreover, $AlLi_2$ and $Al_4Li_9$ appear as lines compounds in both phase diagrams and the maximum temperatures at which both disappear are also very close in both phase diagrams. $Al_2Li_3$ is a line compound in the experimental phase diagram while it is only a line compound up to 500 K in the calculated phase diagram and forms a solid solution at higher temperature.

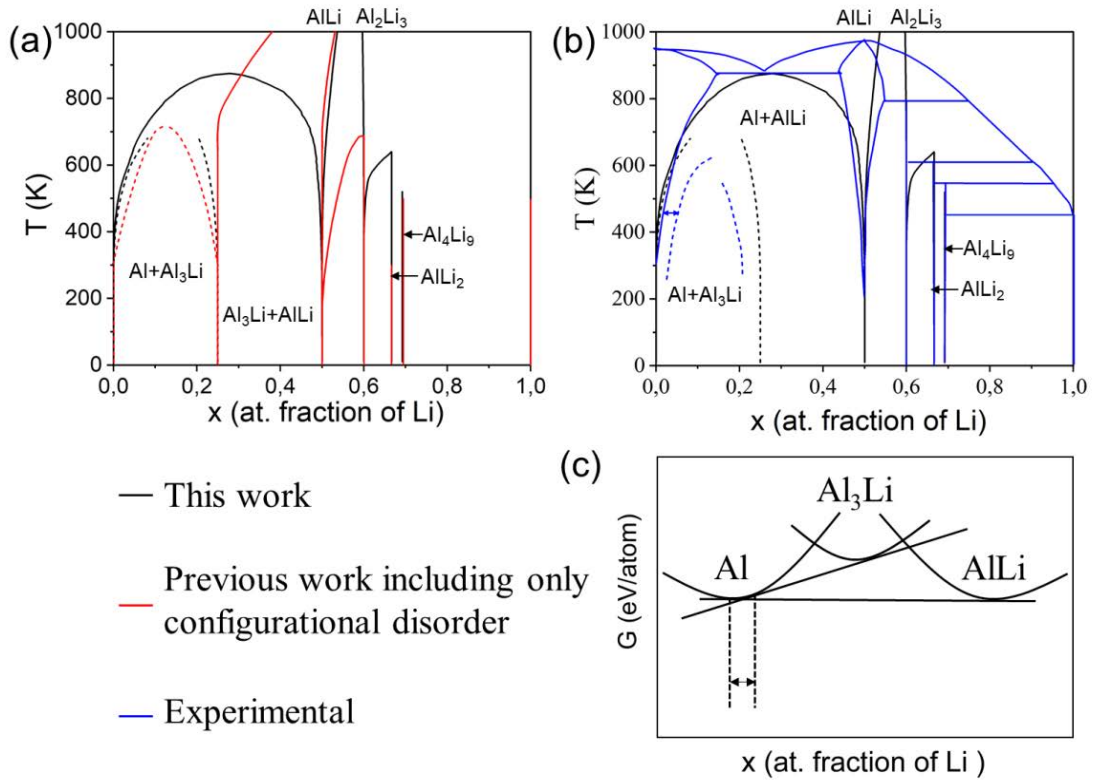

**Fig. 13.** (a) Comparison between Al-Li phase diagrams calculated in this work (black lines) and in a previous investigation (red lines) [29]. The phase diagram calculated in [29] only considered the configurational disorder contribution to the free energy. (b) Comparison between the Al-Li phase diagrams calculated in this work (black lines) and the experimental one (blue lines) [27, 28]. (c) Schematic illustration of the gap between the solvus lines of AlLi and Al₃Li.

Overall, the calculated phase diagram is in very good agreement with the experimental data and the only remarkable difference is the gap between the solvus lines of AlLi and Al₃Li, which is indicated by the arrow in Fig. 13(b). This gap is very small in our calculated phase diagram as compared with the experimental data. A schematic illustration in the formation of gap is displayed in Fig. 13(c), which shows the three curves of the Gibbs free energy of the stable Al and AlLi phases and of metastable Al₃Li at a given temperature. The gap is given by the difference in the composition of Al determined by the common tangent, which is marked by the arrows. A large gap will be formed if the Gibbs free energy of metastable Al₃Li is much higher than the common tangent between the Gibbs free energy of the stable Al and AlLi, (as shown in Fig. 13(c)). However, based on our accurate calculations, the Gibbs free energy of metastable Al₃Li is only slightly higher than the common tangent between the Gibbs free energies of Al and AlLi (Fig. 8), and a large gap cannot be formed. The large gap in the experimental phase diagram is probably an error caused by sluggish kinetics and uncertainties during experiments. Therefore, our calculation provides an accurate solvus line for $Al_3Li$, which is very important for the heat treatment of Al-Li alloy.

## 5. Conclusions

The whole phase diagram of the Al-Li system is predicted from first principles calculations and statistical mechanics taking into account the influence of the configurational and vibrational entropy. To this end, the formation enthalpy of different configurations with either fcc or bcc lattices at different temperatures was accurately predicted by means of cluster expansion formalisms that were fitted from density functional theory simulations. The vibrational entropic contribution of each configuration was determined from the bond length *vs.* bond stiffness relationships for each type of bond (either Al-Al, Li-Li and Al-Li) for each lattice that were obtained from full phonon calculations of various compounds. Thus, the thermodynamic grand potential and the Gibbs free energy of the different phases in the convex hull were obtained as a function of temperature by means of Monte Carlo simulations and the phase diagram was determined.

The calculated phase diagram was in excellent agreement with the experimental data and provides accurate predictions of the stable ($AlLi$, $Al_2Li_3$, $AlLi_2$, $Al_4Li_9$) and metastable ($Al_3Li$) phases, of the phase boundaries between them and of the maximum stability temperature of $AlLi_2$ and $Al_4Li_9$ which are line compounds. In addition, the calculated phase diagram shows that $Al_2Li_3$ is a line compound only up to 500 K and forms a solid solution at higher temperature. It also indicates that gap between the solvus lines of $AlLi$ and $Al_3Li$ is narrower than that found in the accepted experimental phase diagram, an information that is important to design precipitation high treatments for these alloys. Finally, the large influence of the vibrational entropic contribution to assess the metastability of $Al_3Li$, the phase boundaries and the maximum stability temperature of different phases is revealed by comparison with predictions of the phase diagram that only considered the configurational entropy contribution. Overall, the methodology presented in this paper shows that accurate phase diagrams of alloys of technological interest can be predicted from first principles calculations.

## Acknowledgements

This investigation was supported by the European Union's Horizon 2020 research and innovation programme through a Marie Sklodowska-Curie Individual Fellowship (Grant Agreement 893883). Computer resources and technical assistance provided by the Centro de Supercomputación y Visualización de Madrid (CeSViMa) and by the Spanish Supercomputing Network (project FI-2021-3-6) are gratefully acknowledged. Finally, use of the computational resources of the Center for Nanoscale Materials, an Office of Science user facility, supported by the U.S. Department of Energy, Office of Science, Office of Basic Energy Sciences, under Contract No. DE-AC02-06CH11357, is also gratefully acknowledged. Wei Shao acknowledges the support from the China Scholarship Council.

## Data availability

The computational data generated in this investigation can be obtained upon request to the corresponding author.

## References

[1] National Science and technology Council, Materials genome initiative for global competitiveness, Washington DC, America, (2011).

[2] T.M. Pollock, A. Van der Ven, The evolving landscape for alloy design, MRS Bull. 44 (2019) 238-246.

[3] A.K. Mallik, Computer calculations of phase diagrams, Bull. Mater. Sci. 8 (1986) 107-121.

[4] N. Ponweiser, C.L. Lengauer, K.W. Richter, Re-investigation of phase equilibria in the system Al-Cu and structural analysis of the high-temperature phase $\eta_1$-Al$_{1-\delta}$Cu,

Intermetallics 19 (2011) 1737-1746.

[5] O. Zobac, A. Kroupa, A. Zemanova, K.W. Ritcher, Experimental description of the Al-Cu binary phase diagram, Metall. Mater. Trans. A 50 (2019) 3805-3815.

[6] A. Van de Walle, G. Ceder, Automating first-principles phase diagram calculations, J. Phase Equilibria, 23 (2002) 348-359.

[7] C. Oses, E. Gossett, D. Hicks, F. Rose, M.J. Eric Perim, I. Takeuchi, S. Sanvito, M. Scheffler, Y. Lederer, O. Levy, C. Toher, S. Curtarolo, AFLOW-CHULL: Cloud-oriented platform for autonomous phase stability analysis, J. Chem. Inf. Model. 58(12) (2018) 2477-2490.

[8] F. Reif, Fundamentals of statistical and thermal physics, McGraw-Hill. (1965) pp. 66-70.

[9] S. Liu, E. Martínez, J. LLorca, Prediction of the Al-rich part of the Al-Cu phase diagram using cluster expansion and statistical mechanics, Acta Mater. 195 (2020) 317-326.

[10] B. Puchala, A. Van der Ven, Thermodynamics of the Zr-O system from first-principles calculations, Phys. Rev. B 88 (2013) 094108.

[11] A.R. Natarajan, A. Van der Ven, First-principles investigation of phase stability in the Mg-Sc binary alloy, Phys. Rev. B 95 (2017) 214107.

[12] R. Chinnappan, B.K. Panigrahi, A. Van de Walle, First-principles study of phase equilibrium in Ti-V, Ti-Nb, and Ti-Ta alloys, CALPHAD, 54 (2016) 125-133.

[13] Non-metals: thermal phonons. University of Cambridge Teaching and Learning Packages Library, (2020).

[14] P.K. Pathria, P.D. Beale, Statistical mechanics (3 ed.). India: Elsevier. (2011) pp. 201.

[15] V. Ozoliņš, M. Asta, Large vibrational effects upon calculated phase boundaries in Al-Sc, Phys. Rev. Lett. 86 (2001) 448-451.

[16] A. Woźniakowski, J. Deniszczyk, Stability phase diagram of the Ir-Pt solid solution-numerical modelling from first principles, J. Medical Internet. Res. 22 (2013) 265-269.

[17] J.W. Doak, C. Wolverton, V. Ozoliņš, Vibrational contributions to the phase

stability of PbS-PbTe alloys, Phys. Rev. B 92 (2015) 174306.

[18] W. Chen, G. Xu, I. Martin-Bragado, Y. Cui, Non-empirical phase equilibria in the Cr-Mo system: A combination of first-principles calculations, cluster expansion and Monte Carlo simulations, Solid State Sci. 41 (2015) 19-24.

[19] L. Monacelli, R. Bianco, M. Cherubini, M. Calandra, I. Errea, F. Mauri, The stochastic self-consistent harmonic approximation: calculating vibrational properties of materials with full quantum and anharmonic effects, J. Phys.: Condens. Matter. 33 (2021) 363001.

[20] A. Togo, I. Tanaka, First principles phonon calculations in materials science, Scr. Mater. 108 (2015) 1-5.

[21] T. Kamencek, S. Wieser, H. Kojima, N. Bedoya-Martinez, J.P. Durholt, R. Schmid, E. Zojer, Evaluating computational shortcuts in supercell based phonon calculations of molecular crystals: the instructive case of naphthalene, J. Chem. Theory Comput. 16(4) (2020) 2716-2735.

[22] A. van de Walle, G. Ceder, First-principles computation of the vibrational entropy of ordered and disordered $Pd_3V$, Phys. Rev. B 61 (2000) 5972-5978.

[23] M.H.F. Sluiter, M. Weinert, Y. Kawazoe, Force constants for substitutional alloys, Phys. Rev. B 59 (1999) 4100-4111.

[24] I.M. Robertson, Phonon dispersion curves for ordered, partially-ordered and disordered iron-aluminium alloys, J. Phys. Condens. Matter. 3 (1991) 8181-8194.

[25] S.L. Shang, Y. Wang, D.E. Kim, C.L. Zacherl, Y. Du, Z.K. Liu, Strucutral, vibrational, and thermodynamic properties of ordered and disordered $Ni_{1-x}Pt_x$ alloys from first-principles calculations, Phys. Rev. B 83 (2011) 144204.

[26] C. Ravi, B.K. Panigrahi, M.C. Valsakumar, First-principle calculation of phase equilibrium of V-Nb, V-Ta, and Nb-Ta alloys, Phys. Rev. B 85 (2012) 054202.

[27] F. Gayle, J. Vander Sande, A. J. McAlister, The Al-Li (AluminumLithium) system, Bull. Alloy Phase Diagrams, 5 (1984) 19-20.

[28] B. Hallstedt, O. Kim, Thermodynamic assessment of the Al-Li System, Int. J. Mater. Res. 98 (2007) 961-969.

[29] S. Liu, G. Esteban-Manzanares, J. LLorca, First principles prediction of the Al-Li

phase diagram, Metall. Mater. Trans. A 52 (2021) 4675-4690.

[30] J.Z. Liu, G. Ghosh, A. van de Walle, M. Asta, Transferable force-constant modeling of vibrational thermodynamic properties in fcc-based Al-TM (TM = Ti , Zr, Hf) alloys, Phys. Rev. B. 75 (2007) 104117.

[31] P. Giannozzi, S. Baroni, N. Bonini, M. Calandra, R. Car, C. Cavazzoni, D. Ceresoli, G.L Chiarotti, M. Cococcioni, I. Dabo, QUANTUM ESPRESSO: a modular and open-source software project for quantum simulations of materials, J. Phys.: Condens. Mattter 21 (2009) 395502.

[32] J.P. Perdew, K. Burke, M. Ernzerhof, Generalized gradient approximation made simple, Phys. Rev. Lett. 78 (1997) 1396-1396.

[33] H.J. Monkhorst, J.D. Pack, Special points for Brillouin-zone integration, Phys. Rev. B, 13 (1976) 5188.

[34] E.J. Wu, G. Ceder, Using bond-length-dependent transferable force constants to predict vibrational entropies in Au-Cu, Au-Pd, and Cu-Pd alloys, Phys. Rev. B 67 (2003) 134103.

[35] O. Adjaoud, G. Steinle-Neumann, B.P. Burton, A. van de Walle, First-principles phase diagram calculations for the HfC-TiC, ZrC-TiC, and HfC-ZrC solid solution, Phys. Rev. B 80 (2009) 134112.

[36] A. Manzoor, D.S. Aidhy, Predicting vibrational entropy of fcc solids uniquely from bond chemistry using machine learning, Materialia 12 (2020) 100804.

[37] S.Q. Wei, M.Y. Chou, Ab initio calculation of force constants and full phonon dispersions, Phys. Rev. Lett. 69 (1992) 2799.

[38] H. Liu, I. Papadimitriou, F.X. Lin, J. LLorca, Precipitation during high temperature aging of Al-Cu alloys: a multiscale analysis based on first principles calculations, Acta Mater. 167 (2019) 121-135.

[39] B. Montanari, N.M. Harrison, Lattice dynamics of $TiO_2$ rutile: influence of gradient corrections in density functional calculations, Chem. Phys. Lett. 364 (2002) 528.

[40] W. Kang, D. Cheng, C.L. Fu, B.C. Zhou. First-principles investigation of the phase stability and early stages of precipitation in Mg-Sn alloys. Physical Review Materials

4 (2020) 013606.

[41] A. van de Walle. Multicomponent multisublattice alloys, nonconfigurational entropy and other additions to the Alloy Theoretic Automated Toolkit. Calphad 33 (2009) 266-278.

[42] D. Lee, G. Sim, K.J. Zhao, J.J. Vlassak, Kinetic role of carbon in solid-state synthesis of zirconium diboride using nanolaminates: nanocalorimetry experiments and first principles calculations, Nano Lett. 15 (2015) 8266-8270.

[43] S. Kadkhodaei, A. van de Walle, First-principles calculations of thermal properties of the mechanically unstable phases of the PtTi and NiTi shape memory alloys, Acta Mater. 147 (2018) 296-303.

[44] G.D. Garbulsky, G. Ceder, Effect of lattice vibrations on the ordering tendencies in substitutional binary alloys, Phys. Rev. B, 49 (1994) 6327-6330.

[45] A. van de Walle, M. Asta, G. Ceder, The alloy theoretic automated toolkit: a user guide, Calphad. 26 (2002) 539-553.

[46] A.P. Lyubartsev, A.A. Martsinovski, S.V. Shevkunov, P.N. Vorontsov-Velyaminov, New approach to Monte Carlo calculation of the free energy: method of expanded ensembles, J. Chem. Phys. 96 (1992) 1776-1783.

[47] G. Ceder, A derivation of the Ising model for the computation of phase diagram, Comp. Mater. Sci. 1 (1993) 144-150.

[48] C.B. Gopal, A. van de Walle, Self-driven lattice-model Monte Carlo simulations of alloy thermodynamic properties and phase diagrams, Modelling Simul. Mater. Sci. Eng. 10 (2002) 521-538.

[49] K. Ueno, Y. Shibuta, Semi-grand canonical Monte Carlo simulation for derivation of thermodynamic properties of binary alloy, IOP Conf. Ser. 529 (2019) 012037.

[50] H.T. Xue, F.L. Tang, X.K. Li, F.C. Wan, W.J. Lu, Z.Y. Rui, Y.D. Feng, Phase equilibrium of a $CuInSe_2$-$CuInS_2$ pseudobinary system studied by combined first-principles calculations and cluster expansion Monte Carlo simulations, Mater. Sci. Semicond. Process. 25 (2014) 251-257.

[51] B.P. Burton, S. Demers, A. van de Walle, First principles phase diagram calculations for the wurtzite-structure quasibinary systems SiC-AlN, SiC-GaN and SiC-

InN, J. Appl. Phys. 110 (2011) 023507.